\documentclass[10pt, prd, twocolumn, showpacs]{revtex4}

\usepackage{amsmath}
\usepackage{amssymb}
\usepackage{graphicx}
\usepackage{bm} 

\newcommand{\al}[0]{\alpha} 
\newcommand{\be}[0]{\beta}

\newcommand{\dd}[0]{\partial}

\begin{document}

\title{Entropy of thin shells in a (2+1)-dimensional
asymptotically AdS spacetime
and the BTZ black hole limit}
\author{Jos\'{e} P. S. Lemos} \email{joselemos@ist.utl.pt}
\author{Gon\c{c}alo M. Quinta} \email{goncalo.quinta@ist.utl.pt}
\affiliation{Centro Multidisciplinar de Astrof\'{\i}sica - CENTRA,
Departamento de F\'{\i}sica, Instituto Superior T\'ecnico - IST,
Universidade de Lisboa - UL, Avenida Rovisco Pais 1,
1049-001 Lisboa, Portugal\,\,}

\begin{abstract}
The thermodynamic equilibrium states of a static thin ring shell in a
(2+1)-dimensional spacetime with a negative cosmological constant are
analyzed. Inside the ring, the spacetime is pure anti-de Sitter (AdS),
whereas outside it is a Ba\~nados-Teitelbom-Zanelli (BTZ) spacetime
and thus asymptotically AdS. The first law of thermodynamics applied
to the thin shell, plus one equation of state for the shell's pressure
and another for its temperature, leads to a shell's entropy, which is a
function of its gravitational radius alone.  A simple example for this
gravitational entropy, namely, a power law in the gravitational radius,
is given.  The equations of thermodynamic stability are analyzed,
resulting in certain allowed regions for the parameters entering the
problem.  When the Hawking temperature is set on the shell and the
shell is pushed up to its own gravitational radius, there is a finite
quantum backreaction that does not destroy the shell. One then finds
that the entropy of the shell at the shell's gravitational radius is
given by the Bekenstein-Hawking entropy.

\end{abstract}

\maketitle



\newpage

\section{Introduction}

Due to the long-range interaction of the gravitational field,
gravitating systems have important and interesting thermodynamic
properties, such as negative specific heat, making the systems
unstable with consequent gravitational collapse or energy loss through
evaporation. This happens both in Newtonian gravitation and in
general relativity.  A well-known instance of this fact is given by
the black hole system, whose thermodynamic properties were understood
by Bekenstein \cite{bekenstein} and put on a firm basis by Hawking, by
discovering that through quantum effects it radiates at a definite
temperature \cite{hawking}.  Refinement of the study of black hole
thermodynamics appeared in many guises, in particular by the
introduction of a formalism useful for studying general relativistic
systems in a canonical ensemble \cite{york1,brownyork}.

Another gravitating system in general relativity prone to a
thermodynamic study is a thin shell and the spacetime it generates.
Spurred by the interest in black hole thermodynamics, some studies
have analyzed the thermodynamics of thin shells 
in black hole spacetimes \cite{martyork}, or of pure thin shells
in 3+1 spacetimes,
notably in \cite{Mart}, where several thermodynamic quantities of thin
shells are discussed and a stability analysis of them is performed.
Other studies on the thermodynamics of thin shells are
\cite{daviesfordpage,hiscock}.  For related studies of thermodynamics
of gravitating matter, especially quasiblack holes, i.e., stars on the
verge of becoming a black hole, see
\cite{lemoszaslavski1,lemoszaslavskii2}.  All of 
these works are in the
usual 3+1 dimensions.

Now, in many senses, it is interesting to reduce the spatial dimension
by 1 and study general relativity in 2+1 dimensions.  This plays an
important role in the understanding of systems in curved spacetime, as
the decrease in dimensionality with respect to the usual 3+1
spacetime reduces the degrees of freedom to a few. This leaves
possible complications aside and keeps the essential physical
features.  The interest in (2+1)-dimensional general relativity
underwent a boost after a black hole solution was found in spacetimes
with negative cosmological constant, i.e., spacetimes with an anti-de
Sitter (AdS) background \cite{BTZ,BHTZ}.  This (2+1)-dimensional black
hole, the Ba\~nados-Teitelbom-Zanelli (BTZ) black hole, belongs to a
family of solutions, which, depending on the parameters of the
solution, includes the BTZ black holes themselves, positive mass naked
singularities, the AdS spacetime, and negative mass naked
singularities.  The BTZ black hole, the most important solution in the
family, is a black hole solution in its simplest form.  The
singularity it hides is not a curvature singularity, but 
rather is a much
milder topological singularity, akin to the conical singularities
\cite{BTZ,BHTZ}. 

In the realm of thermodynamics and its connection to the quantum
world, the BTZ black hole has a Bekenstein-Hawking entropy $S_{\rm BH}
= \frac14\, 
\frac{A_{\rm h}}{l_{\rm p}}$, where $A_{\rm h}$ 
is the horizon area, in 2+1 dimensions a circumference,
$A_{\rm h}=2\,\pi\,r_+$, $r_+$ is the horizon radius, and $l_{\rm p}$ is the
Planck length given by $l_{\rm p}=G_3\,\hbar$, $G_3$ being the
three-dimensional gravitational constant and $\hbar$ Planck's constant,
and a Hawking temperature given by $T_{\rm H} = \frac{l_{\rm p}}{2 \pi
G_3\,l^2}\,r_+$ \cite{BTZ} (we put $k_{\rm B}=1$ and $c=1$).  The
thermodynamic and entropy properties of the BTZ black hole have been
further explored in, e.g., \cite{brocreighman,zasla21,lem,cadoni}.

In 2+1 dimensions, as in the 3+1 case,
it is also interesting to study the thermodynamics
of self-gravitating thin shells, since the BTZ black hole can form
from the gravitational collapse of such thin shells
\cite{crisostomoolea,mannoh,ortizryan}
which, when static, can be stable or unstable 
according to their intrinsic parameters 
\cite{eiroasimeone}. In 2+1 dimensions, the
thermodynamics of thin shells in spacetimes with zero cosmological
constant has been studied \cite{lemosquinta1}.  Motivated in part by
this study \cite{lemosquinta1} and also from the fact that in 2+1
dimensions in general relativity with a cosmological constant there
are BTZ
black holes with interesting thermodynamic properties, we want
to study in this article the 
thermodynamics of static thin matter shells in 2+1 dimensions.
In particular, we intend to find
the shell's entropy 
and analyze their thermodynamic stability, as well as
scrutinize the limit when the radius of the shell $R$ goes into its
own gravitational radius. To find the spacetime solution we employ
the junction conditions formalism for a thin shell \cite{Israel},
where in this 2+1 case the shell is simply a ring. One can then
determine the pressure and the mass density of the shell in order for
it to be static in a spacetime with a negative cosmological constant,
in which the interior to the shell is pure AdS and the exterior is
asymptotically AdS.  Employing the first law of thermodynamics and
using the formalism for the usual thermodynamic systems \cite{callen},
which was developed by Martinez to apply to thin matter
shell systems in 3+1 general relativity \cite{Mart} 
(see, also, \cite{york1,brownyork}), one finds then the entropy
for these gravitating systems, the desired thermodynamic properties,
and the quasiblack hole limit.

The paper is organized as follows.  In Sec.~\ref{thinsh}, we compute
the components of the extrinsic curvature of the ring shell that leads
to the shell's linear density and pressure.  We also discuss the 
no-trapped-surface condition and the dominant energy condition.  In
Sec.~\ref{thermo} we review the thermodynamics of the shell.  We use
the entropy representation and assume that the state variables are the
proper mass of the shell and its perimeter, or radius. We then use the
first law of thermodynamics for a one-dimensional system to display
the integrability and the stability conditions of the thermodynamic
system.  In Sec.~\ref{eqsos} we present the equation of state for the
pressure in terms of the proper mass and radius of the shell, and we
derive the equation of state for the temperature of the shell as a
function of the state variables. In
Sec.~\ref{entro} we use the previously obtained integrability
conditions for the first law of thermodynamics in order to simplify
the entropy differential of the shell. This allows us to obtain an
expression for the entropy up to an arbitrary function of the
gravitational radius.  In Sec.~\ref{eqstate} we consider a
phenomenological expression for the arbitrary function, consisting in
a power law of the gravitational radius, which allows us to obtain an
explicit expression for the entropy.  We then analyze the
thermodynamic stability of the system by calculating the permitted
intervals of the free parameters in order for the shell to remain
thermodynamically stable. In Sec.~\ref{btz}, the arbitrary function
will be equated to the inverse Hawking temperature, and it will be
found that the Bekenstein-Hawking entropy of the BTZ black hole
naturally arises when the shell is pushed up to its gravitational
radius. Finally, in Sec.~\ref{conc} we draw some conclusions.

\section{The thin shell spacetime}
\label{thinsh}

Einstein's equation in 2+1 dimensions is written as
\begin{equation}
G_{\al\be}-\Lambda g_{\al\be}=8\pi G_3
T_{\al\be}\,, 
\label{ein} 
\end{equation} 
where $G_{\al\be}$ is the Einstein
tensor, $\Lambda$ is the cosmological constant, $g_{\al\be}$ is the
spacetime metric, $8\pi G_3$ is the coupling with $G_3$ being the
gravitational constant in 2+1 dimensions, and $T_{\al\be}$ is the
energy-momentum tensor.  We keep units where the velocity of light is
$c=1$, and thus $G_3$ has units of the inverse of mass. Greek indices
are spacetime indices and run as $\alpha,\beta=0,1,2$, with 0 being
the time index. Since we want to work in an AdS
background where the cosmological constant is negative, we define
the AdS length $l$ through the equation
\begin{equation}
-\Lambda=\frac{1}{l^2}\,.
\label{lads} 
\end{equation}

We now consider a one-dimensional timelike shell, i.e., a ring, 
with radius $R$ in a
(2+1)-dimensional spacetime. The ring divides spacetime into two
parts, an inner region $\mathcal{V}_-$ and an outer region
$\mathcal{V}_+$.  To find the corresponding spacetime solution, we
follow \cite{Israel}.

In the inner region $\mathcal{V}_-$ ($r\leq R$), 
inside the ring, we consider a
spherically symmetric AdS metric, with cosmological
length $l$, given by
\begin{align}\label{LEI}
ds_-^2  & = g_{\al\be}^- dx^\alpha dx^\beta =\nonumber\\&
-\frac{r^2}{l^2}\,dt^2 +   
\frac{dr^2}
{\frac{r^2}{l^2}}
+ r^2\, d\phi^2\,,\quad r\leq R\,,
\end{align}
where polar coordinates $x^{\alpha-}=(t,r,\phi)$ are used.
In the outer region $\mathcal{V}_+$ ($r\geq R$),
outside the shell, the spacetime is described by the
BTZ line element
\begin{align}\label{LEO}
ds_+^2 & = g_{\al\be}^+ dx^\alpha dx^\beta =
-\left(\frac{r^2}{l^2} - 8G_3m\right)\,dt^2 +   
\nonumber\\& \hspace{27mm}
\frac{dr^2}
{\left(\frac{r^2}{l^2} - 8G_3 m\right)} + r^2 \,d\phi^2\,,
\quad r\geq R\,,
\end{align}
written also in
polar $x^{\alpha+}=(t,r,\phi)$ coordinates. Here, $m$ is a
constant which is interpreted as  
the Arnowitt-Deser-Misner (ADM) mass, or energy.
At $r\to \infty$ the spacetime is asymptotically AdS.
On the hypersurface itself, the induced metric 
$h_{ab}$ yields the line element
\begin{equation}
ds_{\Sigma}^2 = h_{ab} dy^a dy^b =
-d\tau^2 + R^2(\tau) d\phi^2,
\end{equation}
where we have chosen $y^a=(\tau,\phi)$ as coordinates on the 
shell and where we have adopted the convention to use latin indexes 
for the components on the hypersurface. 
The shell ring is at radius $R=R(\tau)$, and 
the parametric equations of the ring hypersurface for both the
$\mathcal{V}_-$ and $\mathcal{V}_+$ are $r=R(\tau)$ and $t = T(\tau)$.
The induced metric $h_{ab}$
is written 
in terms of the metrics $g^{\pm}_{\al\be}$ as
\begin{equation}
h^{\pm}_{ab} = g^{\pm}_{\al\be} \, e^{\al}_{\pm}{}_a \,
e^{\be}_{\pm}{}_b
\end{equation}
where $e^{\al}_{\pm}{}_a$ are 
tangent vectors to the hypersurface viewed 
from each side of it.

The formalism employed in \cite{Israel} uses two conditions in order
to assure the smoothness of the metric across the hypersurface.
These are
the junction conditions. The first junction condition states that
\begin{equation}
[h_{ab}]=0\,,
\end{equation}
where the parentheses symbolize the jump in the quantity across the 
hypersurface, here the induced metric. 
This condition leads to the relation
\begin{align} \label{J1}
\left(\frac{r^2}{l^2} -  8G_3m\right) 
\dot{T}^2 - & \frac{\dot{R}^2}{\left(\frac{r^2}{l^2} - 8G_3m \right)} 
\nonumber \\
&= \frac{r^2}{l^2}\dot{T}^2 - 
\left(\frac{r^2}{l^2}\right)^{-1} \dot{R}^2 = 1\,,
\end{align}
where a dot denotes differentiation with respect to $\tau$. 
The
second junction condition makes use of the extrinsic curvature
$K^{a}{}_{b}$ defined as
\begin{equation}
K^a_{\pm}{}_{b} = \nabla_{\be} n_{\al} \, e^{\al}_{\pm}{}_c 
\, e^{\be}_{\pm}{}_b \, h^{ca}_{\pm}\,
\end{equation}
where $\nabla_{\be}$ denotes the covariant derivative
and $n_\alpha$ is the normal to the 
shell. When the jump in
this quantity is non-null, there exists a thin matter shell with
stress-energy tensor $S^{a}{}_{b}$ given by
\begin{equation}
S^{a}{}_{b}=-\frac{1}{8 \pi G_3}
\left([K^{a}{}_{b}]-[K]h^{a}{}_{b}\right)\,,
\end{equation}
where $K = h^{b}{}_{a} K^{a}{}_{b}$. 
For the line elements (\ref{LEI})
and (\ref{LEO}), and using Eq.~(\ref{J1}), one can compute the
nonzero components of $K^{a}{}_{b}$. They are
\begin{align}
K^{\tau}_{+}{}_{\tau} &= \frac{\frac{R}{l^2}+\ddot{R}}{\sqrt{-8G_3
m+\frac{R^2}{l^2}+\dot{R}^2}}\,, \\
K^{\tau}_{-}{}_{\tau} &= \frac{\frac{R}{l^2}+\ddot{R}}{\sqrt{
\frac{R^2}{l^2}+\dot{R}^2}}\,, \\
K^{\phi}_{+}{}_{\phi} &= \frac{1}{R}\sqrt{-8G_3 m+\frac{R^2}{l^2}
+\dot{R}^2}\,, \\
K^{\phi}_{-}{}_{\phi} &= \frac{1}{R}\sqrt{\frac{R^2}{l^2}+\dot{R}^2}\,.
\end{align}
Imposing that the shell is static, i.e., $\dot{R}=0$ and $\ddot{R}=0$,
one finds the non-null components of the stress-energy tensor
for a static shell, 
\begin{align}
S^{\tau}{}_{\tau} & = \frac{
\sqrt{-8G_3m + \frac{R^2}{l^2}}-\frac{R}{l}}{8 \pi
G_3R} \label{S1} \\
S^{\phi}{}_{\phi} & = \frac{1}{8 \pi G_3}\frac{R}{l^2}
\bigg(\frac{1}{\sqrt{-8G_3m 
+ \frac{R^2}{l^2}}} - \frac{1}{\frac{R}{l}}\bigg). 
\label{S2}
\end{align}

If, in addition, we consider the shell to be 
made of a fluid
with  linear energy density
$\lambda$ and pressure $p$, 
the
stress-energy tensor will  have the form
\begin{equation}
S^{a}{}_{b} = (\lambda +p) u^a u_b + p
h^{a}{}_{b}\,,
\label{perffluid}
\end{equation}
where 
$u^a$ is the 3-velocity of a shell element.  
Thus, for such a fluid, we find 
\begin{equation}
S^{\tau}{}_{\tau} =
-\lambda\,,
\label{lam}
\end{equation}
\begin{equation}
S^{\phi}{}_{\phi} = p
\label{press}\,.
\end{equation}
Note that for a one-dimensional fluid with linear energy density and
pressure in a (2+1)-dimensional spacetime, there are only two possible
degrees of freedom to characterize the system. Therefore, the
stress-energy tensor
(\ref{perffluid}) 
is the most general one that one can consider in this setting,
and it is thus seen to be the stress-energy tensor for
a perfect fluid.
Equations (\ref{S1})-(\ref{S2}) together with
Eqs.~(\ref{lam})-(\ref{press})
yield
\begin{align}
\lambda & = \frac{1}{8\pi G_3\, R}
\left({\frac{R}{l}-
\sqrt{-8G_3m + \frac{R^2}{l^2}}}\right)\,,
\label{lambda1}\, \\
p & = \frac{1}{8 \pi G_3} \frac{R}{l^2}
\left(\frac{1}{\sqrt{-8G_3m + \frac{R^2}{l^2}}} - 
\frac{1}{\frac{R}{l}}\right). 
\label{m2}
\end{align} 
Note that $m=0$ means no shell, i.e., 
$\lambda=0$ and $p=0$.

Now, from Eq.~(\ref{LEO}), one finds that 
the gravitational radius $r_+$ of the shell is given by 
\begin{equation}
r_+ = \sqrt{8G_3 m}\, l\,.
\label{gravr}
\end{equation}
It is useful to define a variable $k$ as
\begin{equation}
\label{1L} 
k\equiv
\sqrt{1-\frac{r_+^2}{R^2}}\,.
\end{equation} 
Then Eqs.~(\ref{lambda1})-(\ref{m2}) can be rewritten as 
\begin{equation}
\lambda = \frac{1}{8\pi G_3 l}\left(1-k\right)
\label{mlambda}\,, 
\end{equation} 
\begin{equation}
p = \frac{1}{8 \pi G_3 l}\left(\frac{1}{k} - 
1\right)\,.
\label{pressurefinal}
\end{equation}
Note that when $\lambda=0$ and $p=0$, Eqs.~(\ref{mlambda})
and (\ref{pressurefinal}) [or, if one prefers, Eqs.~(\ref{lambda1})
and (\ref{m2}] give $m=0$, which from 
Eq.~(\ref{LEO}) implies that the outside 
spacetime is pure AdS. Since the inside is also pure AdS there is
no shell in this case, only AdS spacetime.

Having treated the static problem and having 
found $\lambda$ and $p$, there are  
mechanical constraints on the 
shell that should be imposed. 
One constraint is that the shell
must be outside any trapped surface, so that
the spacetime defined by Eqs.~(\ref{LEI})-(\ref{LEO}) 
makes sense.
Imposing that there are no trapped surfaces gives
\begin{equation}
R\geq r_+\,,
\label{notrap}
\end{equation}
i.e., the shell is outside its own gravitational radius.
One can also see what the energy conditions 
lead to. The weak energy condition is automatically satisfied as
we impose $\lambda$ and $p$ non-negative.
On the other hand the dominant energy condition 
$p \leq \lambda$ is equivalent to the relation
$
k^2 - \frac{2+\frac{l^2}{R^2}}{\sqrt{1+\frac{l^2}{R^2}}}
\,k +1 \leq 0
$.
This is satisfied for 
$
\frac{1}{\sqrt{1+\frac{l^2}{R^2}}} \leq k \leq
\sqrt{1+\frac{l^2}{R^2}}
\label{dom22}
$.
The right inequality is trivially obeyed, the left inequality leads
to 
\begin{equation}
R\geq\frac{1}{\sqrt{1-\frac{r_+^2}{l^2}}}\,r_+\,,
\label{dom3}
\end{equation}
which is the equation the shell must obey 
in order that the dominant energy
condition holds. It is new.  It is more stringent than the 
no-trapped-surface 
condition Eq.~(\ref{notrap}).  The condition (\ref{dom3}) is
plausible on physical grounds.  Indeed, for large $l$, the 
spacetime is weakly AdS, and so 
it is an almost flat spacetime, which in three dimensions means no,
or negligible,
gravity. The condition Eq.~(\ref{dom3})
gives then that 
$R\buildrel>\over\sim r_+$, i.e., 
any shell that makes sense, in the sense of Eq.~(\ref{notrap}),
is possible. As $l$ decreases, the spacetime becomes strongly AdS, 
there is strong gravitational
attraction and the shell can satisfy the dominant
energy condition only for sufficient large $R$. When $l=r_+$, the shell
has to have infinite radius in order to obey the dominant energy
condition and for even smaller $l$ there is no shell that obeys the
dominant energy condition.
There are also stability conditions as Eiroa and Simeone have shown
\cite{eiroasimeone}. Although not explicitly shown in 
\cite{eiroasimeone}, presumably the radius $R=R(r_+,l)$ at which the 
shell becomes unstable is slightly larger than 
the $R$ given in Eq.~(\ref{dom3}).

\section{Thermodynamics
and stability conditions for the thin shell: Generics}
\label{thermo}

Now, we assume that the shell is a hot shell, 
i.e., it possesses 
a temperature $T$ as measured locally
and has
an entropy $S$.

In the entropy representation, as stated in \cite{callen}, 
the entropy $S$ of a system is given in terms of 
the state independent variables. 
Following \cite{callen}, when 
$S$ is known,
the thermodynamical system is known. 
We consider as 
the natural state
independent variables the proper local mass $M$
of the shell, and its 
size, here denoted by the perimeter of the 
ring shell $A$. Thus, for the shell,
\begin{equation}
S = S(M,A)\,.
\label{entropyfundamental}
\end{equation}
Thus, when 
$S$ in 
Eq.~(\ref{entropyfundamental}) is known, 
the thermodynamical properties
of the system follow. 

Using these variables, the first law 
of thermodynamics can be written as 
\begin{equation}\label{1LT}
TdS = dM + p \,dA\,,
\end{equation}
where $T$ and $p$ are the
temperature and the pressure conjugate 
to  $A$. In order to find $S$, one has 
to know the equations of state for these quantities,
i.e., 
\begin{equation}\label{eqsstate1}
p=p(M,A)\,,
\end{equation}
and
\begin{equation}\label{eqsstate2}
\beta=\beta(M,A)\,,
\end{equation}
where $\beta=1/T$ is the inverse temperature.

Given Eq.~(\ref{1LT}), one can find its 
integrability condition for the differential of the
entropy. It is given by
\begin{equation}
\label{int1}
\left(\frac{\dd \be}{\dd A}\right)_M = 
\left(\frac{\dd \be p}{\dd M}\right)_A\,.
\end{equation}

There is then the possibility of studying the local intrinsic 
stability of the shell at a thermodynamical level, 
which is guaranteed as long as the inequalities
\begin{equation}\label{C1}
\left(\frac{\dd^2 S}{\dd M^2}\right)_A \leq 0\,,
\end{equation}
\begin{equation}\label{C2}
\left(\frac{\dd^2 S}{\dd A^2}\right)_M \leq 0\,,
\end{equation}
\begin{equation}\label{C3}
\left(\frac{\dd^2 S}{\dd M^2}\right)\left(\frac{\dd^2 S}{\dd
A^2}\right) - \left(\frac{\dd^2 S}{\dd M \dd A}\right)^2 \geq 0\,,
\end{equation} 
are satisfied. For a derivation of this
type of equations see \cite{callen}.

\section{The two equations of state: equation for the pressure
and equation for the temperature}
\label{eqsos}

\subsection{The two independent 
thermodynamic variables}

In order to find the entropy one needs to 
have the equations of state $p=p(M,A)$ and 
$\beta=\beta(M,A)$, see
Eqs.~(\ref{eqsstate1})-(\ref{eqsstate2}), 
with $M$ and $A$ being
the independent variables.

Note, however, that 
\begin{equation}
A=2 \pi R\,,
\label{peri}
\end{equation}
so that the perimeter 
$A$ and the radius $R$ can be swapped at will
as the independent variables.
The definition of the shell's rest mass $M$ is 
\begin{equation}
M = 2 \pi
R \lambda\,,
\label{massdef}
\end{equation} 
where $\lambda$ is given above in Eq.~(\ref{mlambda}),
and so 
\begin{equation}
M = \frac{R}{4 G_3 l}\left(1-k\right)
\label{m}\,.
\end{equation}

We should now put some of the basic quantities, $m$, $r_+$, and 
$k$, in terms of $M$ and $R$. 
Equation (\ref{m}) together with Eqs.~(\ref{gravr}) and (\ref{1L})
implies that 
the ADM mass
$m$ is given in terms of the shell's proper mass $M$
and radius $R$ by 
\begin{equation}\label{M}
m(M,R) = 
\frac{-2G^2_3 M^2 + G_3 M \frac{R}{l}}{G_3}\,,
\end{equation}
so that when there is no shell, i.e., $M=0$, one 
has that the ADM mass of the spacetime is zero, 
$m=0$.
Also now Eq.~(\ref{gravr}) should be written as 
\begin{equation}
r_+(M,R) = \sqrt{8G_3 m(M,R)}\, l\,,
\label{gravrold}
\end{equation}
where $m(M,R)$ is given in Eq.~(\ref{M}), 
and $k$ should be also seen as $k=k(M,R)$, i.e.,
\begin{equation}
\label{1Lold} 
k(M,R)\equiv
\sqrt{1-\frac{r_+^2(M,R)}{R^2}}\,,
\end{equation} 
where $r_+(M,R)$ is given in Eq.~(\ref{gravrold}).

\subsection{The pressure equation of state}

With this rationale in mind, and following the notation
of Eq.~(\ref{eqsstate1}), we 
write the equation for the pressure 
(\ref{m2}) in the form
\begin{equation}
p(M,R) = \frac{1}{8 \pi G_3 l}\left(\frac{1}{k(M,R)} - 1\right)\,.
\label{pressurefinalmr}
\end{equation}
It is clear that Eq.~(\ref{pressurefinal})
yields the equation we aimed for 
as an
equation of state for the shell.
This equation is an exclusive
effect of the gravitational
equations, and the junction 
conditions on the ring, and it does not depend
on the essence of the fields of 
matter composing the shell. In brief,
it is compulsory that the matter fields
obey this equation of state 
so that mechanical equilibrium is maintained.

\subsection{The temperature equation of state}

Now we turn to the other equation of state, Eq.~(\ref{eqsstate2}),
the equation for $p(M,R)$.
Inserting Eq.~(\ref{pressurefinalmr}) 
and the differential of Eq.~(\ref{m}) into
the first law (\ref{1LT}), using 
Eq.~(\ref{peri}), and 
changing variables from $(M,A)$ to 
$(r_+,R)$ to simplify the 
calculations, where $r_+$ is the gravitational radius
of the ring given in Eq.~(\ref{gravrold}),
we obtain
\begin{equation}
\label{1kk} 
dS = \be(r_+,R)\frac{r_+}{4
G_3 \,l\,R\, k} dr_+\,,
\end{equation} 
where now $S$ can be seen as $S=S(r_+,R)$, 
and the same for $\beta$, $\be(r_+,R)\equiv1/T(r_+,R)$.
Equation (\ref{1kk}) is integrable as long as an
appropriate form for $\be$ is given. 
To find this appropriate form for
$\beta$ we use the integrability condition 
Eq.~(\ref{int1}), which upon changing to the 
$(r_+,R)$ variables reads
\begin{equation}\label{de}
\left(\frac{\dd \be}{\dd R}\right)_{r_+} = \frac{\be}{R k^2}\,,
\end{equation}
where $k$ is envisaged now 
as $k=k(r_+,R)$, see Eq.~(\ref{1Lold}).
It can be shown that Eq.~(\ref{de}) 
has the following analytic solution,
\begin{equation}\label{beta}
\be(r_+,R) =\frac{R}{l}\,  k(r_+,R)\, b(r_+)\,,
\end{equation}
where 
$b(r_+)$ is an arbitrary function of the gravitational radius
$r_+$. Note that $b(r_+)$
has units of inverse temperature and can be interpreted as the inverse
of the temperature the shell would possess if located at $R =
\sqrt{l^2+r_+^2}$, as can be seen from Eq.~(\ref{beta}).
Equation  (\ref{beta}) follows 
from the 
integrability condition for 
the entropy and 
is directly related to 
the equivalence principle 
for systems at a temperature different from zero.
It is the Tolman relation for the temperature
in a gravitational system.

Note that $b$ is forced to depend on the state 
variables $(M,R)$ through the specific function 
$r_+(M,R)$. However, 
the 
integrability condition does not 
yield a precise form for $b$.
As stated in \cite{Mart}
(see also \cite{martyork}), this 
is expected on physical grounds as 
the Euclideanized AdS geometry inside 
the ring can be 
identified with any period
in the partition function 
stemming from a path integral approach,
and thus the hot AdS space 
inside the shell can have any temperature,
not fixing $b$ a priori.
Any specific function $b(r_+(M,R))$
must resort to the specificity of the 
matter itself contained in the shell. 
This state of affairs is common in
thermodynamics. In order to find the equation
of state of a gas one can resort to
generic and consistency 
considerations, and find for instance
that the temperature $T$ 
is $T=T(\rho,p,V)$ quite 
generally, where $\rho$, $p$, and $V$
are the density, pressure, and volume of 
the gas, respectively. To have 
then a concrete form for the function
$T=T(\rho,p,V)$, one has to know 
the specificities of the gas, whether it 
is an ideal gas or a Van der Waals gas
with its two specifying constants,
or any other gas, 
see, e.g.,  \cite{callen}.

\section{Entropy of the thin shell}
\label{entro}

We are now in a position to find the entropy $S$
of the thin shell spacetime.
Inserting  Eq.~(\ref{beta}) into Eq.~(\ref{1kk}),  one is led to 
the specific form for the differential of the entropy
\begin{equation}\label{dS1}
dS(r_+) = b(r_+) \frac{r_+ }{4 G_3\,l^2} dr_+\,.
\end{equation} 
Integrating  Eq.~(\ref{dS1})  yields the following entropy, 
$S(r_+) = \frac{1}{4 G_3\,l^2} \, \int b(r_+)\, r_+\,dr_+ + S_0$, 
where $S_0$ is an integration constant.
By
noting that a zero ADM mass shell, i.e., 
$m=0$ or equivalently $r_+ = 0$, should
naturally have zero entropy, for any regular integrand in 
the entropy formula just given we must have, $ S(r_+\to 0) \to 0$,
i.e., $S_0
= 0$. So,
\begin{equation}\label{SM}
S(r_+) = \frac{1}{4 G_3\,l^2} \, \int b(r_+)\, r_+\,dr_+\,.
\end{equation}
In the same way as $b(r_+)$, 
$S$ is also forced to depend on the state 
variables $(M,R)$ through the specific function 
$b(r_+(M,R))$. 
This dependence of $S$ on $r_+(M,R)$
seen in the formula (\ref{SM})  
comes directly 
from the self-gravitating 
nature of the setup.
It is the result of the matching conditions 
(\ref{M}) and (\ref{pressurefinalmr})
which determine the pressure, 
and of the equivalence principle in the form
of the redshift factor of 
the Tolman temperature given in 
(\ref{beta}).
As explained above, a precise
shape for the function $b(r_+(M,R))$
has to emanate from  definite, thermodynamic
or otherwise, 
configurations for the matter fields.

Equation (\ref{SM}) 
opens the possibility of studying the local intrinsic 
stability of the shell at the thermodynamic level, 
which is guaranteed as long as the inequalities
(\ref{C1})-(\ref{C3}) are satisfied.
It also permits us to study the spacetime
thermodynamics in the limit the shell 
approaches its own gravitational radius.

\section{A specific equation of state for the 
temperature of the thin matter shell: Entropy and 
stability}
\label{eqstate}

\subsection{The temperature equation 
and the entropy}

In order to implement the calculation for the entropy,
one must resort to a specific fluid or a specific 
gas and give exactly the function $b(r_+(M,R))$.
One could think of many, and not wanting to
treat here the specificities of the matter, 
we resort to 
the most simple suggestion for $b(r_+)$ as given in
\cite{Mart}, i.e., a
power law equation of the form
\begin{equation}\label{pl}
b(r_+) = 4\,\al\, G_3 \,l^2\frac{r_+^a}{l_{\rm p}^{(2+a)}}\,,
\end{equation}
where $a$ is a free parameter, essentially a number, 
the factors $4
G_3$, $l^2$, and $l_{\rm p}=G_3 \hbar$ appear for dimensional
and useful
reasons, with $l_{\rm p}$ being the Planck's length in a 
three-dimensional
spacetime, and 
$\hbar$ Planck's constant, and $\alpha$ is another free parameter
without units that can be some function of $l/l_{\rm p}$.
For instance, one can choose $\alpha={\bar \alpha}\,\,
\frac{l_{\rm p}^{(2+a)}}{l^{(2+a)}}$, with $\bar \alpha$ a number, 
but many other
choices are possible. Boltzmann's constant is taken as equal to 1.
In order to further justify the choice of the form of $b(r_+)$ in 
Eq.~(\ref{pl}), we recall that in many thermodynamic instances 
one recurs to power law functions, most notably near or 
at a phase transition point, where the temperature 
goes as the power of the density (or the mass) of the fluid
and a power of its specific volume (or the volume itself), 
for instance. These thermodynamic treatments
do not even need to know the details of the 
fine grain constituency of the fluid. Such power laws are
assumed and indeed represent well the fluid behavior.
Here, since the temperature, or what is the same, 
$b$, cannot be any function of $M$ nor any function of $R$, 
and so not a power of $M$ times a power of 
$R$ as one could be led to think from the usual thermodynamic
treatments, 
but has to be a function such that $M$ and $R$ appear 
through $r_+(M,R)$, a natural and simple choice for $b$ is 
that $b(r_+(M,R))$ is a power of $r_+(M,R)$, as 
we have written in Eq.~(\ref{pl}).

Inserting Eq.~(\ref{pl}) into (\ref{SM}) and integrating, we get
\begin{equation}\label{swithout}
S(r_+)
= \frac{\al}{a+2}\left(\frac{r_+}{l_{\rm p}}\right)^{(a+2)}\,,
\end{equation}
valid for any $a$ as long as we consider the case $a=-2$ 
as yielding a logarithmic function, $S(r_+) = \al \ln r_+/l$,
as it should. 
Using this formula for the entropy, one 
is able to analyze the stability conditions 
imposed on the free parameters, and 
despite the fact that the values of the 
parameters $\al$ and $a$ do
not have specific values as long as 
some type of nature of the matter fields is
not prescribed, it is possible to constrain $a$ nonetheless, such that
thermodynamic equilibrium states of the shell are possible.

\subsection{The stability conditions for the specific temperature
ansatz}
\label{stabil1}

As for the stability equations, we
leave the detailed analysis for the 
Appendix 
and present here the main results.
It is then possible to show that
Eqs.~(\ref{C1})-(\ref{C3}) altogether
applied to the temperature and entropy formulas, 
Eqs.~(\ref{pl}) and (\ref{swithout}), respectively,  
imply 
\begin{equation}\label{ain}
-1 \leq a \leq a_c\,,
\end{equation} 
with $a_c$ being a 
root of a polynomial equation having the 
value $a_c=0.255$.

Now, Eq.~(\ref{C1})
implies the inequality
$
\frac{a}{a+1} R^2 \leq r_+^2 \,,
$
which, together with the condition (\ref{notrap}),
that the shell is above or at its own gravitational radius, 
restricts the values for $R$ as
\begin{equation}\label{Cm}
\frac{a}{a+1} R^2 \leq r_+^2 \leq{R^2}
\,.
\end{equation}
Clearly, for $a<-1$, the
left half of the inequality will exceed the right half and thus $a<-1$
is
excluded, that is the reason for the the lower bound in
Eq.~(\ref{ain}).

Turning to Eq.~(\ref{C2}), and combining it with 
Eq.~(\ref{Cm}), one finds that in the 
interval $-1 \leq a \leq0$ 
one has
\begin{equation}\label{Rp}
0 \leq \frac{R}{l} \leq  \left(\frac{(2a+3) + 
\sqrt{\frac{(5a + 9)}{(a+1)}}}{2(-a)}\right)^{\hskip-0.25cm
1/2}\hskip-0.1cm,\,
-1 \leq a \leq 0\,.
\end{equation}

Still, from Eqs.~(\ref{C3}) and (\ref{Cm}),
one finds that in the 
interval $0< a<a_c$, for some critical $a_c$, 
$R/l$ has to obey the equation 
\begin{equation}
a\, \left(1+\frac{R^2}{l^2}\right)^{3/2} +  
\frac{a-1}{\sqrt{a+1}}\frac{R}{l} \leq 0\,,\quad
0< a\leq a_c
\,.
\label{finalstab}
\end{equation}
This means that  in the 
interval $0< a\leq a_c$,
$R/l$ has to be in between two values
$R_{\rm min}(a)$ and $R_{\rm max}(a)$, i.e., 
\begin{equation}
\frac{R_{\rm min}(a)}{l}\leq\frac{R}{l} \leq
\frac{R_{\rm max}(a)}{l}\,,\quad
0< a\leq a_c
\,.
\label{finalstab0}
\end{equation}
The critical value  $a_c$, where the equality 
${R_{\rm max}(a_c)}/{l}={R_{\rm min}(a_c)}/{l}
\equiv R_c/l$
is obtained, is $a_c=0.255$
for which 
$R_c/l={\sqrt 2}/2$.

Finally, through Eq.~(\ref{Cm}), Eqs.~(\ref{Rp})-(\ref{finalstab0})
automatically establish a limit on $r_+$ in order that the shell is
thermodynamically stable.

It is also interesting to note the case where we take the shell to its
gravitational radius, $R=r_+$.  In that situation, and supposing that
the backreaction of whatever kind is small or
negligible, the shell is
thermodynamically stable if $r_+/l$ satisfies the stability conditions
given in Eqs.~(\ref{Rp})-(\ref{finalstab0}), upon substitution of $R/l$
by $r_+/l$ in those.  This result does not take into account the
possible backreaction that might arise due to quantum effects
appearing when the ring is at its own gravitational radius.  In
addition, in this $R=r_+$ limit, although the energy density is
finite, the pressure blows up.

In deriving the stability conditions
Eqs.~(\ref{ain})-(\ref{finalstab0}) the mechanical condition we have
imposed was the no-trapped-surface condition Eq.~(\ref{notrap}).  One
can also rework the stability conditions by imposing the tighter
dominant energy
condition Eq.~(\ref{dom3}) or even the mechanical stability condition
that can be extracted from \cite{eiroasimeone}.
We do not dwell on these conditions as our main interest 
here is on the thermodynamic properties of the shell.

\section{The entropy of the thin shell in the BTZ black hole limit}
\label{btz}

A case of particular interest 
is the marginal case $a=-1$. In this case, 
the inverse temperature $b(r_+)$ of the shell, taken from
Eq.~(\ref{pl}), is
\begin{equation}\label{pl2}
b(r_+) = \frac{4\,\al\, G_3 \,l^2}{l_{\rm p}}  \frac{1}{r_+ }
\,,
\end{equation}
and 
the
entropy of the shell, taken from
Eq.~(\ref{swithout}), has the explicit form
\begin{equation}\label{SBH}
S(r_+) = \al \,\frac{r_+}{l_{\rm p}}\,.
\end{equation}
This is valid for any radius $R$ of the shell
($R\geq r_+$), 
since 
from the integrability condition the
entropy does not depend on the radius 
of the shell $R$.

In particular, if we take the limit 
$R\to r_+$, then the shell hovers at its own 
gravitational radius. One then 
expects that quantum fields are
present \cite{hawking},
and the backreaction will diverge unless one chooses
the matter to be at the Hawking temperature \cite{BTZ}
\begin{equation}
T_{\rm H} = \frac{l_{\rm p}}{2 \pi G_3\,l^2}\,r_+ \,.
\end{equation}
This fixes the function $b=1/T_{\rm H}$ to be
\begin{equation}
b(r_+) = \frac{2 \pi G_3\,l^2}{l_{\rm p}}  \frac{1}{r_+ }\,,
\end{equation}
which means that 
$\alpha=\pi/2$ in Eq.~(\ref{pl2}). 
Then the entropy (\ref{SBH}) of the shell at its 
own gravitational radius is
\begin{equation}
S_{\rm BH} = \frac{\pi}{2} \frac{r_+}{l_{\rm p}}\,.
\end{equation}
The area $A_{\rm h}$ of the horizon,
which
is 
actually a
perimeter
in 2+1 dimensions,
is $A_{\rm h}=2\pi\,r_+$. 
Therefore 
\begin{equation}
S_{\rm BH} = \frac14 \frac{A_{\rm h}}{l_{\rm p}}\,.
\end{equation}
This is precisely the Bekenstein-Hawking entropy of 
the (2+1)-dimensional BTZ black hole 
\cite{BTZ}, now derived from the properties of the
spacetime of the shell of matter
and from the fact that the shell
is at its own gravitational radius.
At $R=r_+$ one has from Eq.~(\ref{1L})
that $k=0$, and so from Eqs.~(\ref{mlambda}) and 
(\ref{pressurefinal}) one finds
that 
$\lambda  = \frac{1}{8\pi\,G_3\,l}$
(i.e., $M  = \frac{1}{4\,G_3}\frac{r_+}{l}$)
and $p= \frac{1}{8\pi\,G_3\,l}\,\frac{1}{k}\to\infty$,
characteristic of 
certain quasiblack holes,
objects which also 
yield the Bekenstein-Hawking entropy
\cite{lemoszaslavski1,lemoszaslavskii2}.
Indeed, the shell at its own gravitational
radius is a quasiblack hole. 

\section{Conclusions} 
\label{conc}

In this paper we have considered the thermodynamics and entropy of a
(2+1)-dimensional shell, a ring, in an AdS spacetime. Inside the ring
the spacetime is given by the AdS metric, characterized by a negative
cosmological constant $-\Lambda=\frac{1}{l^2}$, and outside it is
given by the BTZ metric, characterized by a mass $m$, by the same
negative cosmological constant $-\Lambda=\frac{1}{l^2}$, and by being
asymptotically AdS. The ring shell at radius $R$ has a mass density
$\lambda$ (or equivalently a mass $M=2\pi R \lambda$) and a pressure
$p$ associated with it required to achieve a static equilibrium.

The first law of thermodynamics implies 
that we need to equations of state:
one for the pressure $p$ of the shell and another for its temperature
$T$.  The pressure $p$ is given in terms of the state variables $M$
and $R$ through the junction conditions.  The temperature $T$, or its
inverse, is found to be a function of the gravitational radius $r_+$
of the system alone.  This $r_+$ is itself a particular known function
of the state variables $M$ and $R$.  The entropy of the shell can then
be found as a function of $r_+$ alone.  To give an example of a hot
shell, inspired in several usual thermodynamics systems which have the
temperature as given in power laws of the state variables, we have
chosen the inverse temperature $b$ to be proportional to a power law
in $r_+$, $r_+^a$, for some number $a$.  The computation of the
specific form of the entropy led to an analysis of the parameter
regions for which the ring is thermodynamically
stable. We have found $a$
must be in the range $-1\leq a\leq 0.255$.

In the case $a=-1$, the shell can be chosen to have a Hawking type
temperature at the outset. One can then tune the temperature to be
precisely the Hawking temperature, including all numerical factors,
and push the shell up to its gravitational radius, since at this
temperature there is a finite backreaction at the horizon that does
not destroy the solution.  The entropy found is then the
Bekenstein-Hawking entropy as it is appropriate for a quasiblack hole.

\begin{acknowledgments}
We thank Oleg Zaslavskii for conversations.
We thank FCT-Portugal for financial support through Project
No.~PEst-OE/FIS/UI0099/2013.
\end{acknowledgments}

\appendix \section{Analysis of the stability equations}\label{appendixA}

We use here the stability conditions 
Eqs.~(\ref{C1})-(\ref{C3}) in the 
temperature and entropy formulas, 
Eqs.~(\ref{pl}) and (\ref{swithout}), respectively,  
to show the 
results presented in Sec.~\ref{stabil1}.

It is possible to show that
Eq.~(\ref{C1}) implies the inequality
\begin{equation}\label{CC1app}
\frac{a}{a+1} R^2 \leq r_+^2 \,,
\end{equation}
which, together with the condition 
that the shell is above or at its own gravitational radius, i.e.,
$r_+^2 \leq {R^2}$, sets up the
restricted values for $R$ relative to $r_+$, namely,
\begin{equation}\label{Cmapp}
\frac{a}{a+1} R^2 \leq r_+^2 \leq{R^2}
\,.
\end{equation}
For $0\leq a<\infty$ this inequality always holds. If $-1 \leq a < 0$,
the lower limit will assume negative values, but the inequality is
satisfied nonetheless. For $a<-1$, the left half of the inequality
will exceed the right half and thus $a<-1$ is excluded.  Thus,
Eq.~(\ref{Cmapp}) restricts the interval of $a$ to
\begin{equation}\label{ainapp}
-1 \leq a < \infty\,.
\end{equation} 

Turning now to Eq.~(\ref{C2}), it leads to the relation
\begin{align}\label{CCapp}
r_+^2  + & a\,{R^2} \left(1+ \frac{R^2}{l^2}\right) 
\nonumber \\
& \leq a\, {R^2} \sqrt{1+\frac{R^2}{l^2}}\sqrt{
\frac{R^2}{l^2}-\frac{r_+^2}{l^2}
}\,,
\end{align}
which, when used in conjunction with Eq.~(\ref{CC1app}), leads to
\begin{equation}\label{CC2app}
a(a+1) \frac{R^4}{l^4} + (a+1)(2a+3) \frac{R^2}{l^2} + (a+2)^2 \geq 0.
\end{equation}
Depending on $a$, Eq.~(\ref{CC2app}) gives further
that it can only be verified for a certain set of values 
for $\frac{R}{l}$. Indeed, for $-1 \leq a < 0$ one has
\begin{equation}\label{Rpapp}
0 \leq \frac{R}{l} \leq  \left(\frac{(2a+3) + 
\sqrt{\frac{(5a + 9)}{(a+1)}}}{2(-a)}\right)^{\hskip-0.25cm
1/2}\hskip-0.1cm,\,
-1 \leq a \leq 0\,.
\end{equation}
For $0\leq a<\infty$ one has that any $\frac{R}{l}$ satisfies the
inequality (\ref{CC2app}).
Through
Eq.~(\ref{Cmapp}), this
automatically establishes a limit on $r_+$. 

Finally, from Eq.~(\ref{C3}) one obtains the
inequality
\begin{align}\label{CC3app}
& (a+1)\,\frac{r_+^2}{l^2}\left(1+\frac{R^2}{l^2}\right)^{3/2}
 + 
\nonumber \\
& + \left( a\, \frac{R^2}{l^2}-(a+1)\,\frac{r_+^2}{l^2}
 \right)\sqrt{\frac{R^2}{l^2}
-\frac{r_+^2}{l^2} }  \leq 0.
\end{align}
Imposing at the same time the condition Eq.~(\ref{Cmapp}) 
in Eq.~(\ref{CC3app}), we are left with
\begin{equation}\label{CC3appplus}
a\, \left(1+\frac{R^2}{l^2}\right)^{3/2} +  
\frac{a-1}{\sqrt{a+1}}\frac{R}{l} \leq 0\,.
\end{equation}
It is clear that 
for $-1\leq a\leq0$
the inequality (\ref{CC3appplus}) is always satisfied.
It is also clear that there is a critical $a$, $a_c$, 
such that 
for $0< a\leq a_c$, $R/l$ has to be in
between two values,
and for 
$a_c< a<\infty$, 
the inequality (\ref{CC3appplus}) can never be satisfied.
The value of $a_c$ and the corresponding
value $R_c/l$ are found as follows. Consider
$f(R)\equiv a\, \left(1+\frac{R^2}{l^2}\right)^{3/2} +  
\frac{a-1}{\sqrt{a+1}}\frac{R}{l}$. From the form 
of the function $f(R)$, one can check that there is an $a_c$ 
above which the inequality (\ref{CC3appplus}) has no solution. 
To this $a_c$ there is a correspondent $R_c/l$.
Imposing $f(R)=0$ and $df(R)/dR=0$, one is led to
one equation for $a$ and one for $R/l$, namely,
$27\,a^3+23\,a^2+8\,a-4=0$ and $R^2/l^2-1/2=0$. The solutions are 
$a_c=0.255$, up to the third decimal place, 
and $R_c/l={\sqrt 2}/2$.
 
Collecting the results given in this 
Appendix, we have for thermodynamic stability that
the following equations must be satisfied: 
\begin{equation}\label{ainx}
-1 \leq a \leq a_c\,,
\end{equation} 
with  $a_c = 0.255$,
\begin{equation}\label{Cmx}
\frac{a}{a+1} R^2 \leq r_+^2 \leq{R^2}
\,,
\end{equation}
\begin{equation}\label{Rpx}
0 \leq \frac{R}{l} \leq  \left(\frac{(2a+3) + 
\sqrt{\frac{(5a + 9)}{(a+1)}}}{2(-a)}\right)^{\hskip-0.25cm
1/2}\hskip-0.1cm,\,
-1 \leq a \leq 0\,,
\end{equation}
and 
\begin{equation}
a\, \left(1+\frac{R^2}{l^2}\right)^{3/2} +  
\frac{a-1}{\sqrt{a+1}}\frac{R}{l} \leq 0\,,\quad
0< a\leq a_c
\,, 
\label{finalstabxxx}
\end{equation}
i.e., 
\begin{equation}
\frac{R_{\rm min}(a)}{l}\leq\frac{R}{l} \leq
\frac{R_{\rm max}(a)}{l}\,,\quad
0< a\leq a_c
\,.
\label{finalstab0xxx}
\end{equation}
The critical value of $R/l$, namely, 
${R_{\rm max}(a_c)}/{l}={R_{\rm min}(a_c)}/{l}
\equiv R_c/l$,
is obtained for $a_c=0.255$, yielding
$R_c/l={\sqrt 2}/2$.

Equation (\ref{Cmx}), with the help of 
Eqs.~(\ref{ainx})-(\ref{finalstab0xxx}),
automatically establishes
a limit on $r_+$ in order that the shell is
thermodynamically stable.
These are the results presented in Sec.~\ref{stabil1}.

\end{document}